\documentclass[11pt,preprint]{aastex}
\usepackage{amsmath,amssymb,rotating,graphicx,natbib}

\newcommand{\epseri}{$\epsilon$~Eridani\ }

\newcommand{\appropto}{\mathrel{\vcenter{
  \offinterlineskip\halign{\hfil$##$\cr
    \propto\cr\noalign{\kern2pt}\sim\cr\noalign{\kern-2pt}}}}}

\shorttitle{Apocenter Glow}
\begin{document}

\title{Apocenter Glow in Eccentric Debris Disks:\\ Implications for Fomalhaut and $\epsilon$ Eridani}

\author{Margaret Pan\altaffilmark{1,2,3}, Erika R.~Nesvold\altaffilmark{4,5}, Marc J.~Kuchner\altaffilmark{3}}
\altaffiltext{1}{MIT Department of Earth, Atmospheric, and Planetary Sciences, Cambridge, MA 02139}
\altaffiltext{2}{Department of Astronomy and Astrophysics, University of Toronto, Toronto, Ontario M5S~3H4}
\altaffiltext{3}{NASA Goddard Space Flight Center, Exoplanets and Stellar Astrophysics Laboratory, Greenbelt, MD 20771}
\altaffiltext{3}{Department of Applied Physics, University of Maryland Baltimore County, Baltimore, MD 21250}
\altaffiltext{4}{Department of Terrestrial Magnetism, Carnegie Institution of Washington, Washington, DC 20015}

\begin{abstract}
 
Debris disks often take the form of eccentric rings with azimuthal
asymmetries in surface brightness.  Such disks are often described as
showing ``pericenter glow'', an enhancement of the disk brightness in
regions nearest the central star. At long wavelengths, however, the
disk apocenters should appear brighter than their pericenters: in the
long wavelength limit, we find the apocenter/pericenter flux ratio
scales as $1+e$ for disk eccentricity $e$.  We produce new models of
this ``apocenter glow'' to explore its causes and wavelength dependence
and study its potential as a probe of dust grain properties. Based on
our models, we argue that several far-infrared and (sub)millimeter
images of the Fomalhaut and \epseri debris rings obtained with {\it
  Herschel}, JCMT, SHARC II, ALMA, and ATCA should be reinterpreted as
suggestions or examples of apocenter glow. This reinterpretation
yields new constraints on the disks' dust grain properties and size
distributions.

\end{abstract}

\section{Introduction}
\label{sec:introduction}

More and more high-resolution images show that debris disks often take
the form of rings, sometimes narrow, sometimes eccentric.
Well-resolved examples include HR4796 \citep{schneider09, thalmann11,
  lagrange12}, Fomalhaut \citep{stapelfeldt04, kalas05}, HD 181327
\citep{schneider06}, $\zeta^2$ Reticuli \citep{eiroa10}, HD 202628
\citep{krist12}, and HD 115600 \citep{currie15}.  These rings may
indicate the presence of hidden planets, which can clear the central
cavities in the rings and also excite the ring eccentricities via
secular perturbations \citep{roques94, wyatt99, kuchner03, quillen06,
  chiang09, rodigas14, nesvold15}.

At shorter wavelengths, regions of an eccentric disk near pericenter
appear brighter because they receive more flux from the host
star. \citet{wyatt99} named this phenomenon ``pericenter glow'' and
developed a model for an eccentric debris ring interacting with a
single planetary perturber. Their model disk consists of massless
particles whose eccentricities differ only in the direction of the
free eccentricity.  The resulting ring suffices to explain the offset
in the solar zodiacal cloud from the sun and to fit observations of
several debris disks, yielding constraints on the disks' forced
eccentricity and the masses of the hidden planetary perturbers.

However, variations in the disk surface density also affect its
apparent brightness, and in a steady-state disk the density should
peak at apocenter simply because typical orbit velocities are slowest
there. Though modeling done by \citet{wyatt99} predicted a brightness
enhancement at pericenter for HR 4796 at 18.2~$\mu$m, their disk model
showed a 2\% density enhancement at apocenter. Analogous apocentric
density enhancements occur in more recent dynamical models of
eccentric planets interacting with disks \citep[see, for
  example,][]{nesvold13, pearce14}.  Indeed, submillimeter
observations of the very well-observed eccentric Fomalhaut disk
consistently suggest apocentric brightness enhancements.  JCMT images
of Fomalhaut by \citet{holland03} show slight enhancements of the flux
near apocenter; these enhancements are less than the quoted
uncertainty in the photometry, but they appear in both 450~$\mu$m and
850~$\mu$m bands.  When \citet{marsh05} imaged the Fomalhaut disk at
350~$\mu$m using the SHARC II (Submillimeter High Angular Resolution
Camera II) at the Caltech Submillimeter Observatory, they found that
the ring has an apocentric enhancement of approximately 14\% in
integrated column density.  More recently, \citet{ricci12} imaged
Fomalhaut's disk at 7~mm with the ATCA and noted that the lobe of the
disk near apocenter ``...appears to be more extended, showing two
possible asymmetric structures toward east and south.'' The highest
resolution ALMA images of Fomalhaut by \citet{boley12} at 350~GHz
(1~mm) also show enhanced flux near apocenter in the maps corrected
for the single-dish beam.

Here we describe a new model for debris rings that illustrates the
wavelength dependence in the apocenter/pericenter flux ratio due to
the competing effects of azimuthal asymmetries in dust density and
stellar illumination. Our primary interest here is mid-infrared and
longer wavelengths, so we focus on dust particles and planetesimals
large enough to avoid radiation pressure effects and consider only
absorbed and re-radiated, rather than scattered, emission. In Section
\ref{sec:semianalytic}, we begin by describing a semi-analytic model
for estimating the surface density of a steady-state distribution of
collisionless planetesimals and show that the density of a dust ring
varies with longitude and peaks at apocenter. In Section
\ref{sec:smack}, we verify this result for a collisional ring using
SMACK \citep{nesvold13}, a numerical model of debris disk evolution
that incorporates both collisions and dynamics in 3D. Finally in
Section \ref{sec:rerad}, we combine a simple dust reradiation model
with our models of surface density to simulate the brightness of the
Fomalhaut and $\epsilon$ Eridani rings, and demonstrate that the ratio
between pericenter and apocenter flux varies with wavelength. We
summarize our results and discuss the implications for future
observations of eccentric rings in Section \ref{sec:conclusions}.

\section{Semi-Analytic Model}
\label{sec:semianalytic}

The surface brightness of an optically thin debris disk depends on the
underlying spatial distribution of the dust as well as on the dust
particles' absorption, reradiation, and scattering properties.  In
this section, we use basic orbit geometry and a simple Monte Carlo
simulation to estimate the surface density of a eccentric ring of
collisionless particles and its dependence on longitude from
pericenter.

\subsection{Disk Density Calculations}

We first estimate the linear number density, $\ell$, of an eccentric
ring of particles as a function of longitude, $f$, in the
ring. $\ell(f)$ is the most relevant quantity for the many
images of disks that are unresolved in the radial direction. We assume
that the particles form an annulus about the star, and that their
eccentricity, inclination, and semimajor axis distributions are
centered respectively on values $e$, 0, $a$ and have widths $\Delta
e<e\ll 1$, $\Delta i\simeq \Delta e$, $\Delta a\simeq a\Delta
e$. Because we are considering a ring that appears eccentric overall
and has a fractional radial width no larger than $\Delta e$, we must
constrain the longitude of pericenter $\varpi$ to a distribution
centered on $\varpi=0$ with width
\begin{equation}
\Delta\varpi \simeq \sqrt{1-e^2}\,\Delta e/e .
\label{eqn:deltaomega}
\end{equation}
Also, we assume the particles' orbital phases and longitudes of the
ascending node are distributed uniformly.

For a single particle in a stable elliptical orbit given by
\begin{equation}
r=\frac{a(1-e^2)}{1+e\cos f} \;\;\;,
\label{eqn:r}
\end{equation}
the time-averaged linear number density along the orbit is inversely
proportional to the local Keplerian velocity:
$\ell_\mathrm{single}(f)\propto 1/v(f)$ where $v(f)$ is given by
\begin{equation} \label{eqn:velocity}
\begin{split}
v(f) & = \left(\frac{GM_*}{a}\frac{1+2e\cos f+e^2}{1-e^2}\right)^{1/2} \\
 & \simeq v_0\left(1+e\cos f +e^2\left(1-\frac{\cos^2 f}{8}\right)\right) \\
v_0 & = \sqrt{GM_*/a}\;\;\;.
\end{split}
\end{equation}
Here $v_0$ is simply the orbital velocity of a circular orbit
with the same $a$.  We consider only mildly eccentric disks ($e\ll 1$),
so to lowest order in $e$ the linear number density in the disk scales
as
\begin{equation}
  \ell_\mathrm{single}(f) \propto \frac{1}{v(f)}\propto 1-e \cos f.
  \label{eqn:ell}
\end{equation}
In short, because particles orbit faster at pericenter and slower at
apocenter, their number density decreases at pericenter and increases
at apocenter by the same fractional amount $e$. \citet{marsh05} also
calculated $\ell(f)$ by a somewhat different method and found a similar
enhancement at apocenter.

In a disk of such particles, the overall linear density $\ell(f)$ is
the sum of all the individual densities
$\ell_\mathrm{single}(f-\varpi)$. For a disk with particle orbit
elements within the ranges given above, the linear density dependence
on $f$ should remain close to that of $\ell_\mathrm{single}(f)$, that
is,
\begin{equation}
\ell(f)\appropto 1-e \cos f \;\;\;,
\end{equation}
with slight variations due to the finite widths of the distributions
of orbit elements.

\subsection{Monte Carlo Simulations}

To check our scaling argument above, we performed each Monte Carlo
disk simulation by randomly drawing $5\times 10^4$ disk particle
orbits from orbit element distributions as follows. The semimajor
axes, eccentricities, longitudes of pericenter, and inclinations were
drawn from Gaussian distributions with center values and widths
($\sigma$) given by
\begin{itemize}
\item typical eccentricity $e$ and eccentricity width $\Delta e$ chosen for each disk;
\item typical semimajor axis $a=1$ and semimajor axis width $\Delta a=\Delta e$;
\item typical longitude of pericenter $\varpi=0$ and corresponding width $\Delta\varpi$ given by Equation~\ref{eqn:deltaomega};
  \item typical inclination $i=0$ and inclination width $\Delta i=\Delta e$.
\end{itemize}
The particles' longitudes of the ascending node and mean anomalies were chosen from uniform distributions over $[0,2\pi)$.

We then measured the linear density as a function of longitude for
each simulation, an example of which is shown in
Figure~\ref{fig:lindensity}. As expected, $\ell$ varies nearly
sinusoidally with $f$ with amplitude about equal to the eccentricity.


In the above density calculations we assumed the grains are
distributed uniformly along their orbits. In particular, we ignored
azimuthal variations in dust production caused by differences in the
frequency and severity of collisions between larger planetesimals
induced, for example, by perturbations from planets orbiting in the
system \citep[see, for example,][]{nesvold15}. These should be
unimportant in our work here on gravitationally bound particles in a
steady state disk, since any effects of azimuthal variations in their
production rate should smear out within an orbit period\footnote{While
  azimuthal variations in production rate might indeed affect the
  steady state distribution of unbound dust small enough to be
  affected by radiation pressure, we leave studies on their behavior
  to future work}. In the next section we describe a numerical test of
this effect.

\section{Numerical Model with SMACK}
\label{sec:smack}

The semi-analytic model described in Section
\ref{sec:semianalytic} assumes that the given ranges of orbit elements
accurately represent a steady state disk. To test these assumption and
the results of the semi-analytic model, we simulated the collisional
and dynamical evolution of the planetesimals using SMACK
\citep{nesvold13}. SMACK uses an N-body integrator to track the orbits
of superparticles, clouds of parent bodies with a range of sizes. When
two superparticles collide, SMACK updates the superparticles'
trajectories and size distributions to account statistically for
collisions between their planetesimals and tracks the mass of dust
(\textless 1 mm in size) produced in the collision.

We modeled the evolution of a ring of planetesimals in orbit around a
solar-mass star. The planetesimals were represented by 10,000
superparticles of uniform size whose orbital elements were initially
uniformly distributed with the ranges in Table~\ref{tab:initial}. Each
superparticle represented a size distribution of planetesimals with
range and slope given in Table~\ref{tab:initial}. We set the initial
vertical optical depth of the ring to $10^{-4}$. While the semi-analytic model
includes the eccentric ring alone, in order to induce and maintain an
eccentricity in the SMACK model we added a 3 $M_\mathrm{Jupiter}$
planet to the system at semimajor axis 50 AU with eccentricity 0.1,
then evolved the system for $10^7$ yr.

\begin{deluxetable}{lc}
\tablewidth{0pt}
\tablecaption{Initial conditions for the SMACK model. \label{tab:initial}}
\tablehead{\colhead{Parameter	} & \colhead{Initial Disk Values} }
\startdata
	Semimajor Axis ($a$) 		& 50-100 AU\\
	Eccentricity ($e$)				& 0.0-0.2 \\
	Inclination ($i$)				& 0.0-0.1 \\
	Longitude of Ascending Node ($\Omega$)	& 0-$2\pi$ \\
	Argument of pericenter ($\omega$)		& 0-$2\pi$ \\
	Mean Anomaly 				& 0-$2\pi$ \\
        Size distribution index ($q$)& $-3.5$\\
        Planetesimal sizes & 1~mm to 10~cm\\
        Optical depth& $10^{-4}$\\
        Number of superparticles& $10^4$
\enddata
\end{deluxetable}

\begin{figure}
  \centering
  \includegraphics[scale=0.8]{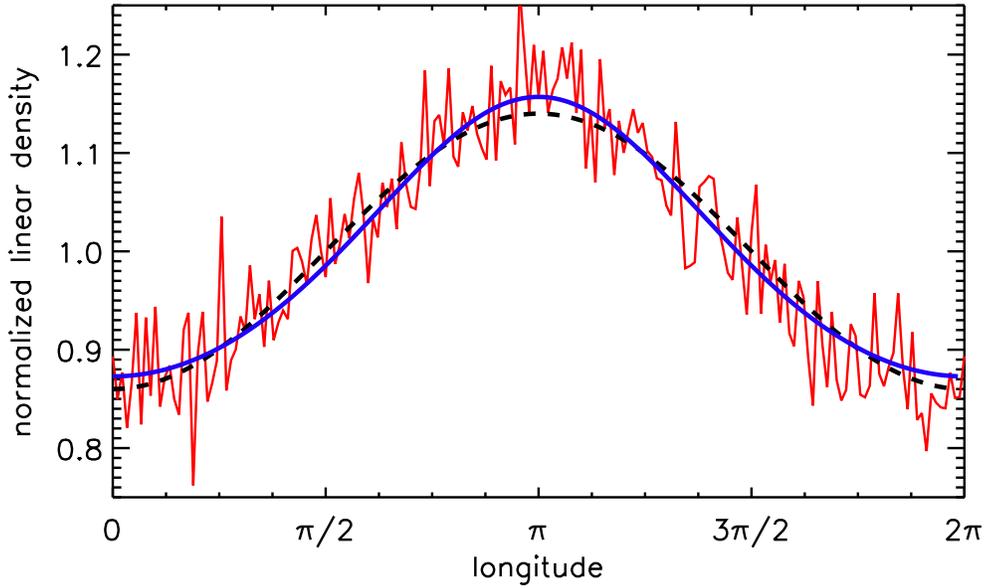}
  \caption{Linear number density $\ell$ as a function of longitude $f$
    in our simulated debris disks. SMACK model data is the thin red
    line; MC model data is the solid blue line; the $1-e\cos f$ we
    predict analytically is the dashed black line. The median
    eccentricity of the SMACK model disk is $e=0.14$ and its
    eccentricity width is $\Delta e=0.07$; the same $e$ and $\Delta e$
    were used in the MC model disk shown here. Both sets of model data
    were normalized to their respective mean values. $e=0.14$ was also
    used for the $1-e\cos f$ curve. While the MC model appears to be a
    very slightly better match to the SMACK model data than the cosine
    curve --- the root mean square deviation between the SMACK and MC
    models is about 2\% smaller than that between the SMACK model and
    the cosine --- all three agree well overall. Note that no fitting
    was performed; the normalized model data are simply overplotted
    along with the cosine.}
  \label{fig:lindensity}
\end{figure}

A snapshot of the simulation at this point shows that most of the
superparticle eccentricities are distributed in a single roughly
symmetric peak with half-width $\Delta e\simeq 0.07$, while perhaps
10\% form a tail extending to larger eccentricities (0.3 to 0.5). The
peak eccentricity value is about 0.12, the median is about 0.14, and
the mean is about 0.15.  This distribution is broadly similar to our
semi-analytic model assumptions except in that $\Delta e\simeq e$
rather than $\Delta e\ll e$. The corresponding normalized disk linear
density is shown in Figure \ref{fig:lindensity} and is roughly
sinusoidal with a peak at apocenter and amplitude about equal to the
median eccentricity. Again, this is broadly consistent with our
semi-analytic and MC model predictions. As we would expect, the
inclusion of collisions in the SMACK model does not strongly affect
the normalized linear density. By definition the eccentric disk has
some steady-state finite eccentricity $e$ that persists regardless of
the collisions; the Keplerian orbit shape required by this $e$ sets
the linear density variation with longitude. Also, because the
optical depth is much less than unity, any non-uniformity in dust
production spreads around the disk much faster than dust can
accumulate at a given longitude.

\section{Dust Reradiation Model}
\label{sec:rerad}

We have used semi-analytic and SMACK modeling to simulate the density
of the dust ring produced by an eccentric parent body ring and showed
that the orbital geometry of the dust produces a peak in surface
density at apocenter. The observed flux from the dust is therefore
subject to two competing effects: this higher number density at
apocenter and the higher temperature at pericenter discussed in detail
by \citet{wyatt99}. We now combine our semi-analytic and SMACK results
with a simple radiative equilibrium model for dust emission to study
the disk emission as a function of longitude at different
wavelengths. We apply this surface brightness model to two
well-observed disks: Fomalhaut and $\epsilon$ Eridani.

\subsection{Dust Model Framework}

We assume a passively heated disk containing dust of sizes $\{s\}$
with size distribution $dN/ds\propto s^{-q}$. We assume the disk is optically thin so that the thermal equilibrium condition for each
grain is
\begin{equation}
  \int d\lambda A(s,\lambda)L_*\frac{\pi s^2}{4\pi r^2}
  = \int d\lambda A(s,\lambda)4\pi s^2\sigma_\mathrm{SB}T^4(s,f)
\end{equation}
where $r=a(1-e^2)/(1+e\cos f)$ is the orbital radius, $T$ is the
dust's effective temperature, and the absorptivity $A(s,\lambda)$ is 1
when $s>\lambda$ and scales as $(s/\lambda)^\beta$ otherwise.  We
solve this iteratively for sizes $s$ over a grid of wavelengths
$\lambda$ including the star's blackbody peak and sum over the sizes
$s$, with weights assigned according to $dN/ds$, to find the flux at
each of the desired wavelengths.

\subsection{Fomalhaut}
\label{sec:fomalhaut}

Because it is so well-observed at wavelengths from UV to mm-wave
radio, the Fomalhaut disk \citep[eccentricity $\sim$0.1,][]{kalas05}
provides an excellent test for our models. As discussed in
Section \ref{sec:introduction}, previous works have consistently found that
Fomalhaut exhibits pericenter glow --- i.e., is brightest at its
southeast limb --- at wavelengths shorter than about 250~$\mu$m
\citep[see, for example,][and references
  therein]{kalas05,acke12}. However, observations at longer submillimeter
wavelengths consistently suggest that the northwest limb appears brighter
than would be expected in a uniform ring \citep{holland03,marsh05,boley12}. 

\begin{figure*}
\includegraphics[height=\textwidth,angle=90]{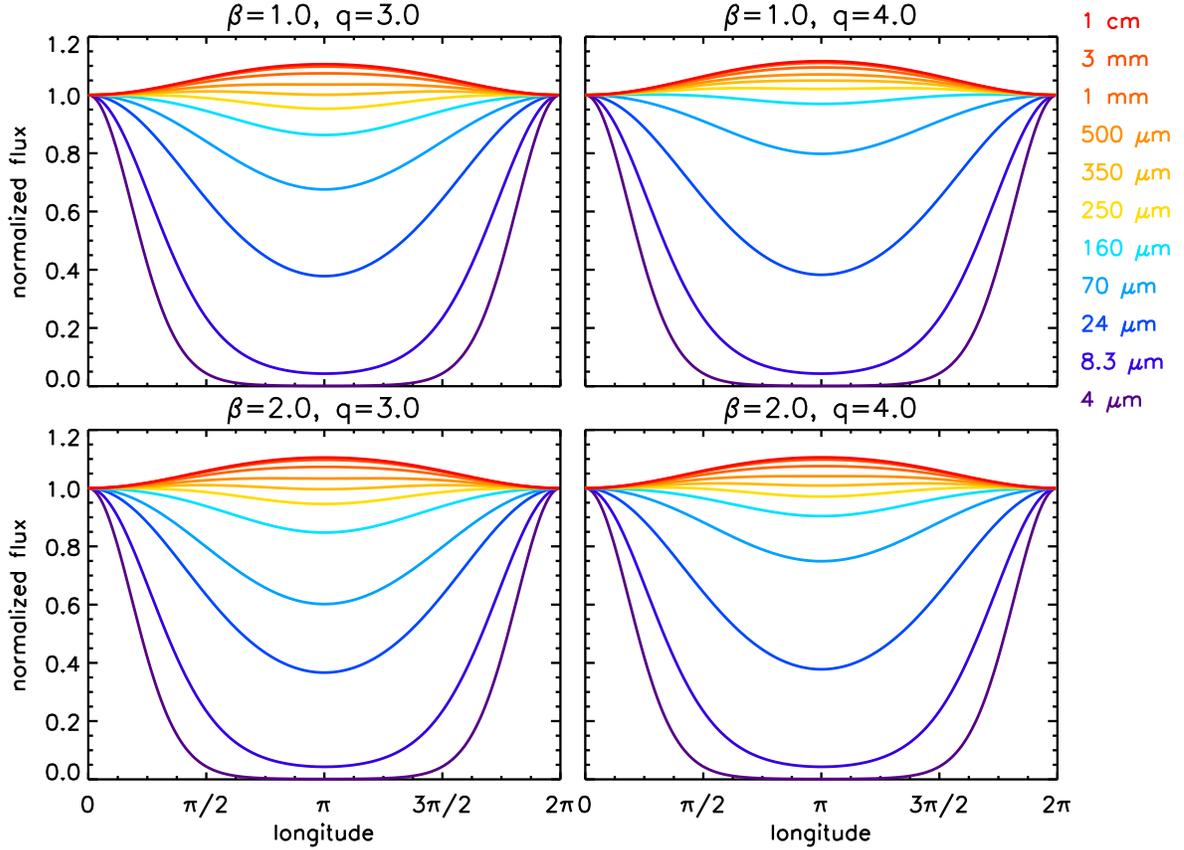}
\caption{Examples of results of our dust reradiation simulations of
  the Fomalhaut disk. Each panel shows the radially integrated disk
  flux as a function of longitude, normalized to pericenter, for 12
  different wavelengths, shown as different-colored curves. The top
  row shows results for absorptivity $\propto(s/\lambda)^1$ and the
  bottom row corresponds to $(s/\lambda)^{2}$. The left and right
  columns show results for $q=3$ and $q=4$ respectively. These
  represent the extremes of the $\lambda$ -- $q$ grid we explored in
  our dust reradiation simulations. Shorter wavelengths demonstrate
  pericenter glow, but longer wavelengths exhibit apocenter glow
  instead. Note the qualitative variation from apocenter to pericenter
  in the 160~$\mu$m flux (light blue curve) across the panels.}
\label{fig:fomalhautmodels}
\end{figure*}

We applied our dust reradiation model to the Fomalhaut system using
stellar temperature $T_*=8590$~K, stellar radius $R_*=1.28\times
10^{11}$~cm, disk semimajor axis $a=133$ AU, radial width $\sim$20~AU,
and eccentricity 0.1. To approximate the results of our semi-analytic
model, we vary the linear mass density of the disk sinusoidally with
longitude with amplitude 0.1 (simulating a disk with eccentricity
$\sim$0.1). We explored a grid of absorptivity laws
$1.0\leq\beta\leq2.0$ and dust size distributions $3\leq q\leq 4$ to
predict the apocenter/pericenter glow from visual to millimeter
wavelengths.  Figure \ref{fig:fomalhautmodels} shows some examples of
our results. Each panel shows the flux as a function of longitude,
integrated over the radial width of the disk and normalized to
pericenter, for 12 different wavelengths shown as different-colored
curves. At the longer wavelengths, the simulated ring exhibits
apocenter glow rather than pericenter glow. The location of the peak
in flux shifts from apocenter (longitude = $\pi$) to pericenter
(longitude = 0) with decreasing wavelength. The shortest wavelength to
show apocenter glow in Figure \ref{fig:fomalhautmodels} is typically
160 or 70~$\mu$m.

\begin{figure}
\begin{center}
  \includegraphics[width=\columnwidth]{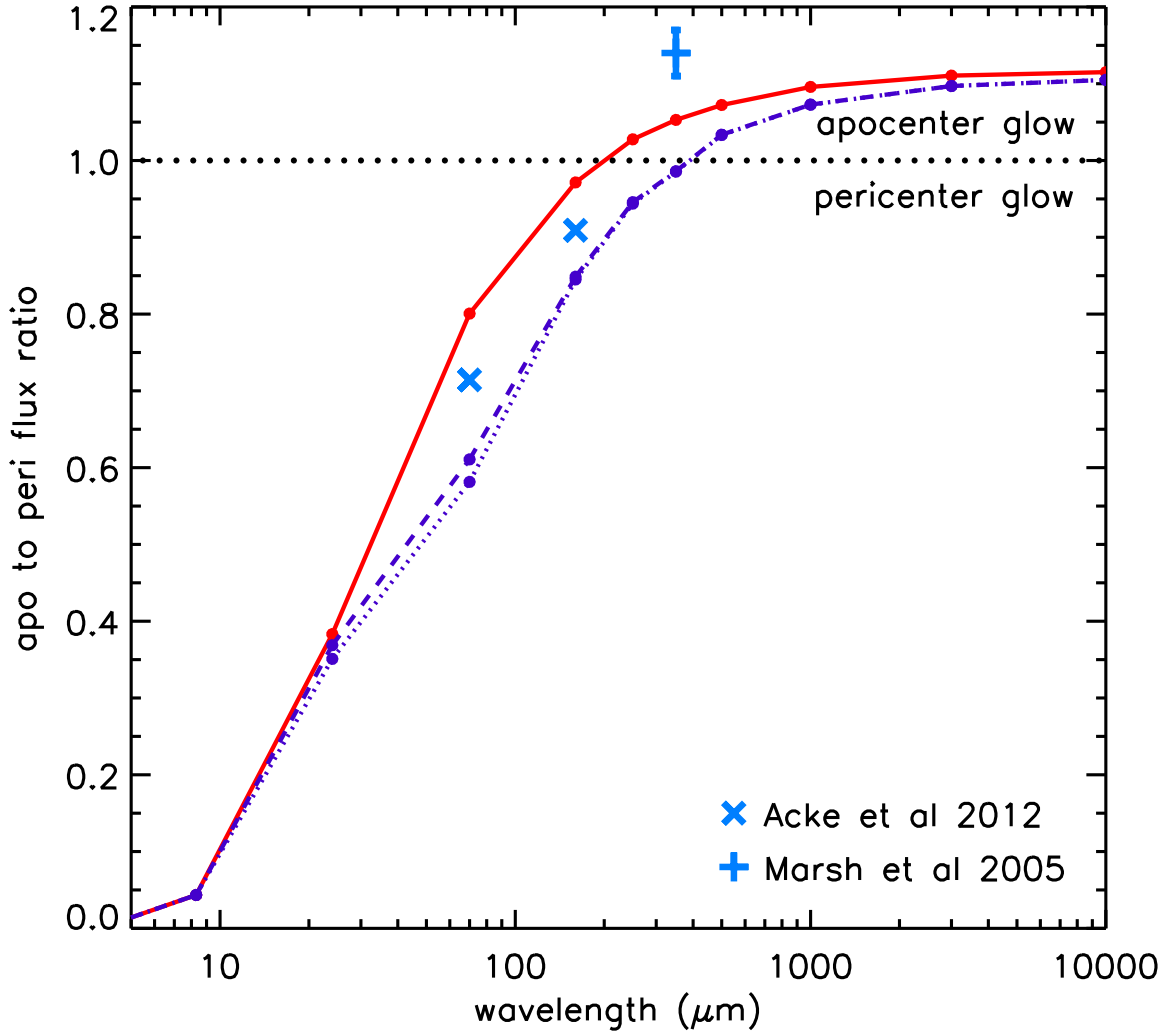}
  \caption{Apocenter to pericenter flux ratios (ratio of the radially
    integrated disk flux at apocenter to that at pericenter) as a
    function of wavelength for our grid of Fomalhaut dust reradiation
    models. The red solid curve follows the maximum flux ratio values,
    which consistently occur at $q=4$, $\beta=1$; the purple
    dashed/dotted curves follow the flux ratios occurring at $q=3$ and
    $\beta=2$ (dashed) or $\beta=3$ (dotted). The $q=3$, $\beta=3$
    flux ratios are the minimum attained on our parameter grid:
    extending our upper bound on $\beta$ from 2 to 3 makes little
    difference in the overall range of model flux ratios. Blue points
    correspond to measured flux ratios from Herschel data at 70 and
    160~$\mu$m \citep{acke12} and from CSO/SHARC II data at 350~$\mu$m
    \citep{marsh05}. The observed results show broad agreement with
    our dust reradiation model output. Uncertainties on the
    \citet{acke12} points are smaller than the plot symbols.}
  \label{fig:fluxampfomalhaut}
\end{center}
\end{figure}

We show a summary view of the apocenter/pericenter flux contrast in
Figure \ref{fig:fluxampfomalhaut}, which displays the range of
apocenter/pericenter flux ratios occurring across our grid of models
as a function of wavelength. Again, the modeled fluxes were integrated over
the radial width of the disk, since azimuthal width variations are not
well resolved in the images. Here, models exhibiting pericenter glow
lie below the black dotted line while models with apocenter glow lie
above it. At short wavelengths, all our models exhibit pericenter
glow, while at long wavelengths, they all exhibit apocenter glow. This
illustrates the competition between higher temperatures at pericenter
discussed in detail by \citet{wyatt99}, which for Fomalhaut affect
smaller particles more strongly and dominate in the shorter-wavelength
emission, and the higher densities at apocenter, whose signature
emerges strongly in the long-wavelength emission. Far-infrared and
submillimeter observations of Fomalhaut appear broadly consistent with
the range of apocenter/pericenter flux ratios our models predict,
though flux ratio values based on ALMA and ATCA observations longward
of 350~$\mu$m are currently not precise enough to include in
Fig.~\ref{fig:fluxampfomalhaut}.

A sufficiently large beam can dilute the effect of apocenter glow or
blur it with background galaxies or other disk features like clumps,
or even disk ansae, which themselves can be limb-brightened if the
disk is not face-on. For example, in their discussion of pericenter
glow in the Keck image of HR~4796 from \citet{telesco00},
\citet{wyatt99} calculate the effects of beam size and study the
disappearance of disk asymmetry as pericenter/apocenter moves away
from the disk ansae. Because the beam size in the Fomalhaut images is
much smaller relative to the disk size than in the \citet{telesco00}
HR 4796 observations, we believe these dilution/confusion effects are
less important for the Fomalhaut data of \citet{acke12} and
\citet{marsh05}\footnote{We refer here to the deconvolved image of
  \citet{marsh05}; the deconvolution method uses knowledge of the PSF
  to extract extra resolution information from strongly overlapped
  individual fields of view.}. Indeed both sets of disk images appear
ring-shaped rather than dumbbell-shaped/double-lobed as in the earlier
HR 4796 images. Also, for Fomalhaut \citet{acke12} measure the disk
major axis to coincide within 1 degree with the line of nodes in the
sky plane, so it is highly unlikely that pericenter orientation masks
the pericenter/apocenter flux ratio.

In these relatively well-resolved images we believe we can reasonably
estimate the uncertainty in the flux ratio due to beam size by
convolving a one-dimensional Gaussian of width equal to the beam
radius with a cosine representing the sinusoidal flux variation with
longitude and quoting the difference in the pre- and post-convolution
amplitudes. In order to compensate here for the apparent increase in
flux at the disk ansae due to Fomalhaut's 66$^\circ$ inclination, we
convolve the cosine with a one-dimensional Gaussian of width equal to
the beam radius divided by $\cos 66^\circ$.  For the {\em Herschel}
data this gives an extra flux uncertainty of $\sim$5\% of the
difference between the pericenter and apocenter fluxes at 70 microns
and $\sim$13\% of this difference at 160 microns, smaller than the
size of the plot symbols. \cite{marsh05} incorporated beam/orientation
effects in their data analysis so we infer they are included in the
quoted uncertainties\footnote{For the deconvolved SHARC II image our
  one-dimensional estimation method gives an uncertainty of $\sim$3\%,
  which is comparable to that of \citet{marsh05}. Nonetheless, in our
  modeling and analysis we simply adopt the result and uncertainty
  reported by \citet{marsh05}; we assume their analysis correctly
  accounted for any effects the deconvolution may have had on the
  relative flux, as necessary.}.

\begin{figure}
  \begin{center}
  \includegraphics[scale=.5]{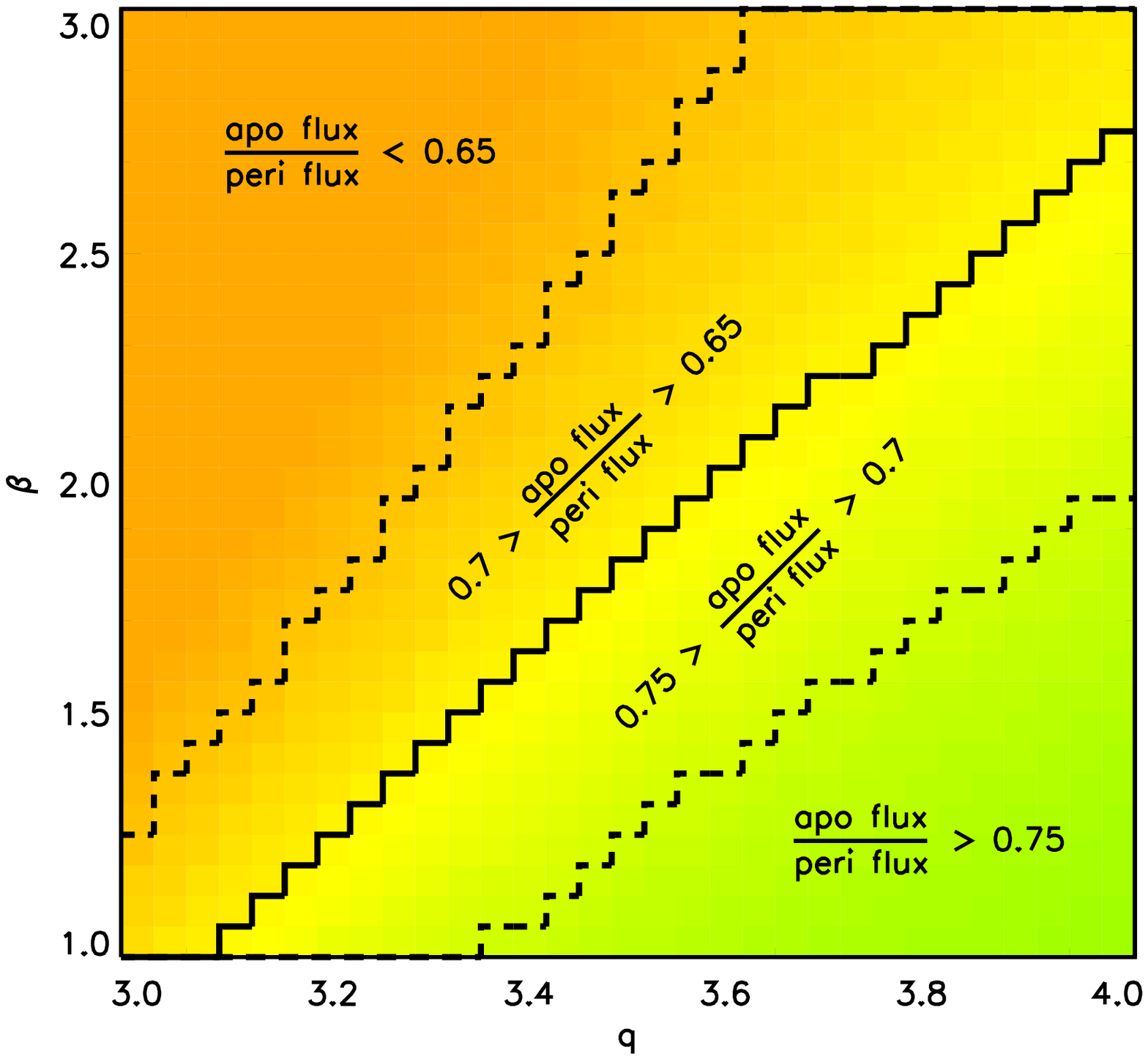}
  \includegraphics[scale=.5]{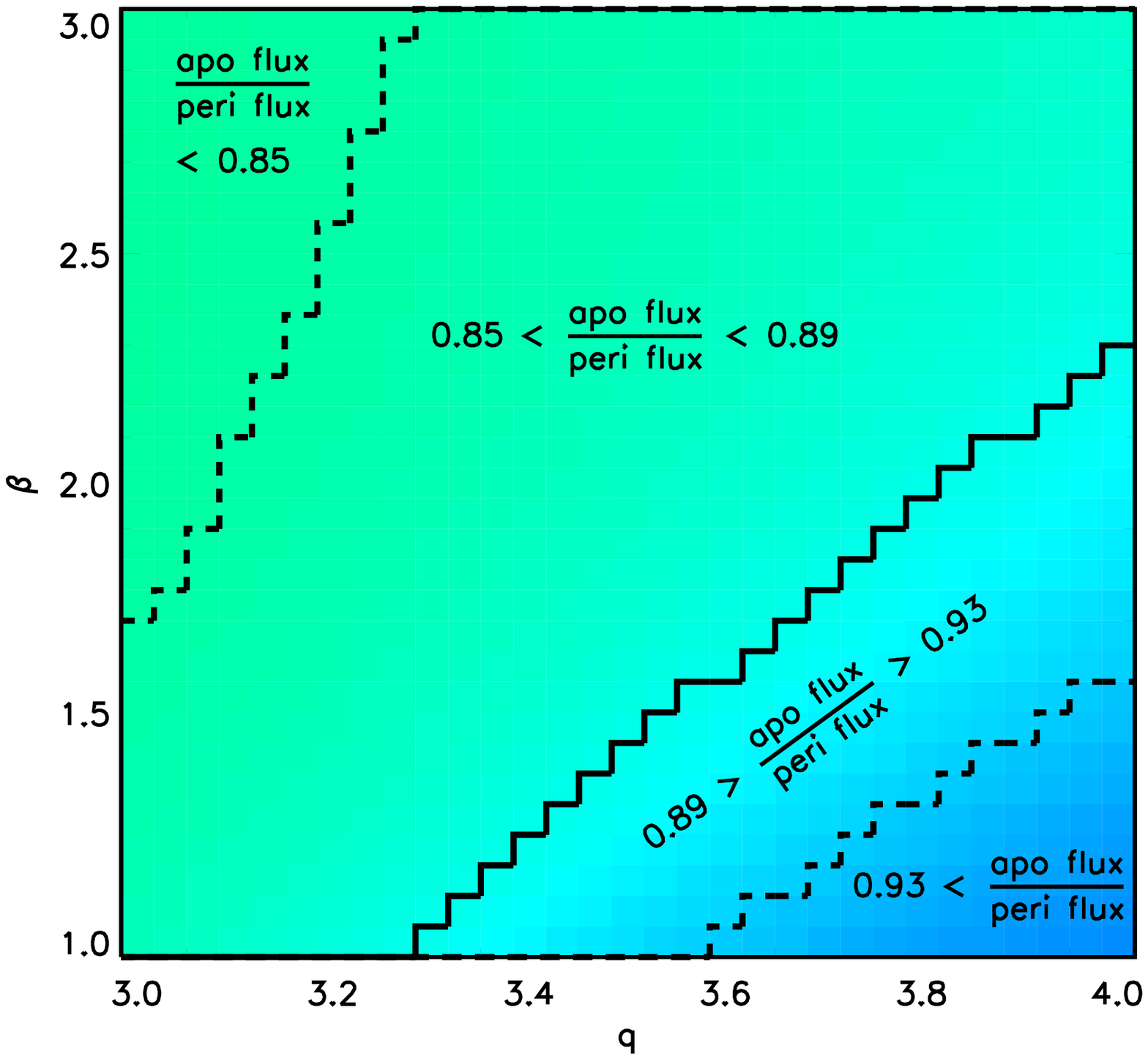}
  \caption{Fractional amplitudes (referenced to the pericenter flux)
    of the 70~$\mu$m (top) and 160~$\mu$m (bottom) fluxes in our grid
    of Fomalhaut dust reradiation models. All values in our parameter range
    produce pericenter glow at these wavelengths. The solid contours
    correspond to 
    apocenter/pericenter flux ratios of 0.7 (top) and 0.89 (bottom), the values extracted from
    Herschel observations by \cite{acke12}. The dashed contours mark nearby values of the flux ratio and give an idea of how the ratio changes as a function of the size distribution slope $q$ and the absorptivity slope $\beta$.}
\end{center}
  \label{fig:fomalhautmodels160}
\end{figure}

Precisely where the observed flux ratios fall among our models can be
a revealing diagnostic of disk properties. Figure
\ref{fig:fomalhautmodels160} shows the pericenter to apocenter flux
ratio at 70 and at 160~$\mu$m for our Fomalhaut models as a function
of $\beta$ and $q$. A {\em Herschel} image of the Fomalhaut disk at
70~$\mu$m shows pericenter glow with an apocenter/pericenter flux
ratio of about 0.7 \citep{acke12}. This value corresponds
approximately to the line $\beta\simeq 1.9q-3.2$. At 160~$\mu$m, the
Fomalhaut ring also shows pericenter glow \citep{acke12}, but the
derived apocenter/pericenter flux ratio is about 0.89, corresponding
approximately to the line $\beta \gtrsim 1.8q - 4.7$.

Observations of Fomalhaut at 7~mm by \citet{ricci12} provide an
additional constraint: $q=3.48\pm 0.14$.  Combining these constraints
with our findings above from the {\em Herschel} 70- and 160-$\mu$m
data and our models yields the range $\beta \simeq 1.4$ to 1.7,
somewhat higher than the $\beta=1.1$ obtained by \citet{dent00} in
their Fomalhaut model. Large values of $\beta$ often point to a lack
of larger, millimeter-sized grains in the grain distribution. For
example, in the outer regions of T~Tauri disks where grain growth has
not yet occurred, $\beta$ is often in the range of 1.7-2
\citep{perez12}. The millimeter dust opacity slope for the ISM yields
$\beta\approx 1.7$ \citep{li01}. However, the Fomalhaut system is too
old for lack of grain growth to be likely. Explaining the Fomalhaut
disk's strong pericenter glow may require additional physics not
included in our models, for example radiation from unbound grains or
planet-disk interactions more complex than those we considered
(e.g.~resonant dynamics). Indeed, \citet{acke12} find that unbound
grains contribute about a quarter of the non-stellar flux in their
70~$\mu$m Fomalhaut system simulations.

Finally, 350~$\mu$m observations by \cite{marsh05} using SHARC II at
the CSO indicate somewhat stronger apocenter glow than our models
predict. Either a very steep size distribution ($q>4$), a very shallow
absorptivity law ($\beta <1$) or, perhaps least unlikely, a larger
eccentricity $e>0.1$ is required in our models to reproduce the
\cite{marsh05} result. In fact \citet{acke12} measure $e=0.125$ and
$e=0.17$ respectively from their 70~$\mu$m and 160~$\mu$m images.

\subsection{\boldmath $\epsilon$ Eridani}

Resolved images of the $\epsilon$ Eridani cold dust disk
\citep[$a\simeq 61$~AU][]{greaves14} were recently obtained at several
submillimeter/millimeter wavelengths with {\em Herschel}, {\em SCUBA},
and {\em MAMBO} \citep{greaves14,lestrade15}. These images also
display azimuthal asymmetry, providing an independent constraint on
our models. Because \epseri \citep[K2V, $T_*\simeq 5084$~K, $R_*\simeq
  5.12\times 10^{10}$~cm,][]{kovtyukh03} is much cooler than Fomalhaut,
none of its associated dust is subject to radiation pressure or
stellar wind blowout \citep{reidemeister11}. Small grains may
potentially provide much of the disk's surface area. Following
\cite{reidemeister11}, we included grains down to 0.1~$\mu$m in our
models of \epseri. As with Fomalhaut, we produced a grid of
models with $3\leq q\leq4$ and $1\leq\beta\leq 3$. For \epseri
we also examined a range of disk eccentricities $0.02\leq e\leq 0.25$,
allowing the amplitude of the surface density variation with longitude
to scale linearly with eccentricity, as demonstrated in our
semi-analytic model.

\begin{figure}
\begin{center}
\includegraphics[scale=.7]{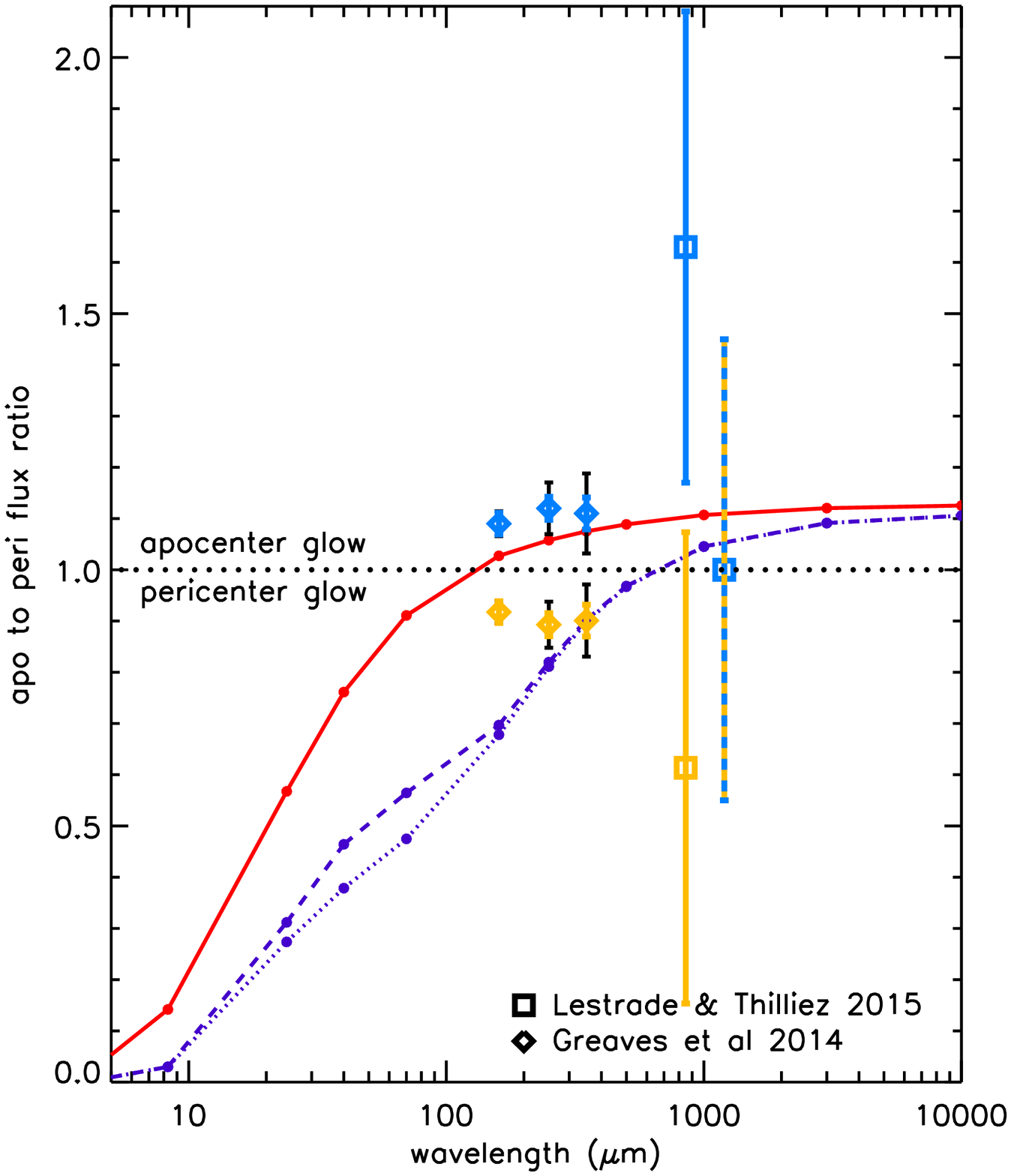}
\caption{Same as Figure~\ref{fig:fluxampfomalhaut}, but for the outer
  disk of $\epsilon$ Eridani \citep[K2V, $T_*=5084$~K,
    $a=61$~AU][]{kovtyukh03,greaves14}. The dust reradiation models
  represented here use the disk eccentricity favored by
  \cite{greaves14}, $e=0.1$. Plotted points correspond to south to
  north flux ratios derived from {\em Herschel}
  \citep[crosses][]{greaves14} and {\em SCUBA} and {\em MAMBO}
  \citep[squares][]{lestrade15} observations. Because the pericenter
  direction has not been independently determined, the measured flux
  ratios may be interpreted as pericenter (orange points) or apocenter
  (blue points) glow. Thin black error bars on the {\em Herschel} data correspond to flux uncertainties due to confusion within the beam. Note that the pericenter glow interpretation of
  the {\em Herschel} measurements at 350~$\mu$m is inconsistent with
  all of our models by about $3\sigma$ (see text for discussion of
  plotted uncertainties): to reproduce that scenario, we would require
  a size distribution significantly shallower than $q=3$ and/or an
  absorptivity slope much steeper than $\beta=3$.}
\label{fig:fluxampepseri}
\end{center}
\end{figure}

Figure \ref{fig:fluxampepseri} shows our modeling results of \epseri
with $e=0.1$, the disk eccentricity favored by \cite{greaves14},
plotted together with the observed south to north flux ratios. As
these fluxes were reported in a variety of formats in the discovery
papers, we converted them as follows. \cite{greaves14} report that
their 160~$\mu$m, 250~$\mu$m, and 350~$\mu$m flux ratios differ from
unity by 3.0, 3.7, and 2.5 times the spread measured in a 9-pixel grid
around each point. Assuming a roughly Gaussian distribution of pixel
fluxes, the spread among 9 pixels would be $\simeq$2$\sigma$ where
$\sigma$ is the Gaussian width parameter. We produced our plotted
uncertainties by assigning an uncertainty of $\pm\sigma$ to the north
and the south fluxes and propagating errors. Although \epseri is
nearly face-on, with an inclination of $\sim$26 degrees, we estimate
using the convolution method described in section~\ref{sec:fomalhaut}
that the large {\em Herschel} beam size relative to the disk size adds
uncertainties of 27\%, 42\%, and 71\% of the pericenter-apocenter flux
difference for 250, 350, and 500 microns
respectively. \cite{lestrade15} report in their Figure~5 the
850~$\mu$m and 1200~$\mu$m fluxes as a function of azimuth. For each
wavelength we took the 5 points closest to due north and due south,
averaged them to get the north and south fluxes, and took the larger
of the spread in the points or the uncertainties plotted by
\cite{lestrade15} as the overall uncertainty in the north/south
fluxes. We then propagated errors to get the uncertainties plotted in
Figure~\ref{fig:fluxampepseri}. These overwhelm the flux uncertainties
due to beam size.

Because the \epseri disk has no independent pericenter direction
determination, the flux asymmetry may be interpreted as either
pericenter or apocenter glow. A pericenter glow interpretation of the
{\em Herschel} 160-, 250-, and 350-$\mu$m observations is consistent
with the range of flux ratios given by our models as long as $e\geq
0.02$. However, plots analogous to those in
Figure~\ref{fig:fomalhautmodels160} indicate that with a pericenter
glow interpretation, the 160~$\mu$m data consistently favors large $q$
(a steep size distribution) and small $\beta$ (a shallower
absorptivity law) while the 350~$\mu$m favors the opposite end of our
parameter range, small $q$ and larger $\beta$.  By contrast, an
apocenter glow interpretation of the {\em Herschel} measurement is
compatible with our models only if $e\gtrsim 0.2$. In this case all
three {\em Herschel} flux ratios are consistent with a steep size
distribution (large $q$) and a shallower absorptivity law (small
$\beta$). While our models consistently predict apocenter glow only
longwards of 850~$\mu$m, the precision of the flux ratios gleaned from
the \citet{lestrade15} data are such that our models appear broadly
consistent with either a pericenter or an apocenter glow interpretation.

\subsection{Warmer disks}

\begin{figure}
\begin{center}
  \includegraphics[width=\columnwidth]{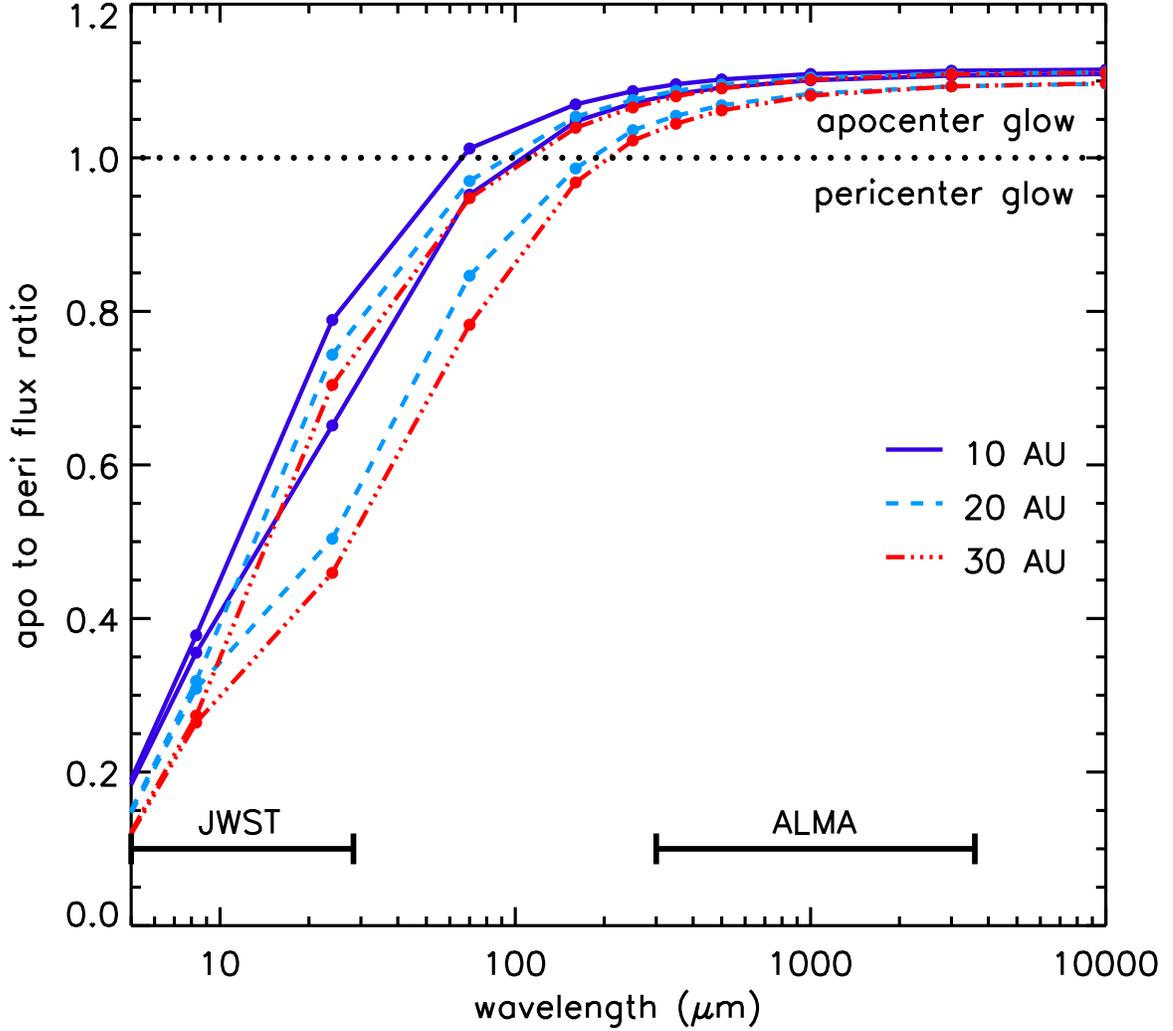}
  \caption{Apocenter/pericenter flux ratios as a function of
    wavelength for model debris disks of sizes 10~AU (solid purple),
    20~AU (dashed blue), and 30 AU (dash-dotted red) around a
    Fomalhaut-like A star. As before, our grid of dust reradiation
    models covered parameter ranges $1\leq\beta\leq 3$, $3\leq q\leq
    4$. The top curve of each color/line style follows the maximum
    flux ratio values in our grid of models, while the bottom curve
    follows the minimum values. The wavelengths at which the greatest
    range of flux ratios occur are 24 to 70 $\mu$m. This suggests that
    observations of pericenter vs. apocenter asymmetries at the longest
    JWST bands are well-placed to constrain $\beta$ and $q$ in
    close-in disks around A stars. }
  \label{fig:fluxampsmalldisk}
\end{center}
\end{figure}

Figure 4 suggests that with its maximum operating wavelength of 28.3
microns, the James Webb Space Telescope (JWST) may not be able to add
much to our understanding of pericenter/apocenter asymmetries in
Fomalhaut: our Fomalhaut models are degenerate in this band.  However,
JWST could play an important role for warmer disks. As a final
application for our dust reradiation model, we studied disks with
semimajor axes of 10, 20, and 30 AU around a Fomalhaut-like A~star.
Figure~\ref{fig:fluxampsmalldisk} shows a summary of the results. Due
to the higher effective temperature of the dust, we expect the
transition wavelength between pericenter and apocenter glow to occur
in the far-infrared rather than the submillimeter bands that are
important for Fomalhaut and $\epsilon$ Eri. The longest JWST MIRI
bands are well-placed to constrain $\beta$ and q in such disks around
A stars by measuring pericenter/apocenter asymmetries. Even barely
resolved observations at 24 to 30~$\mu$m could constrain the size
distribution and eccentricities of those disks.

\section{Summary and Discussion}
\label{sec:conclusions}

Using both semi-analytic and numerical modeling of the azimuthal dust
distribution in an eccentric ring of colliding planetesimals, we have
studied the wavelength dependence of surface brightness variations
using simple assumptions about dust radiative properties and size
distributions.  We argued that several far-infrared and
(sub)millimeter images of Fomalhaut and $\epsilon$ Eridani obtained
with Herschel, JCMT, SHARC II, ALMA, and ATCA
\citep{holland03,marsh05,ricci12,boley12,greaves14} should be
reinterpreted as suggestions or examples of apocenter glow. This
reinterpretation also yields new constraints on the grain properties
and size distributions from the existing data.

Our modeling work also has implications for future observations of debris
disks. The James Webb Space Telescope will be a powerful new source of
debris disk images, observing at 5-28 microns with the MIRI instrument.
Figure~\ref{fig:fluxampsmalldisk} illustrates that this wavelength range is
particularly sensitive to pericenter glow. We predict apocenter/pericenter
flux ratios as small as 0.1 in this range for dust emitting via the highly
temperature-sensitive Wien law.

{\em ALMA}, on the other hand, will operate at wavelengths from 3mm to
400$\mu$m, primarily continuing to detect apocenter glow. {\em ALMA}
images will be especially useful because, as
Figures~\ref{fig:fluxampfomalhaut}, \ref{fig:fluxampepseri},
\ref{fig:fluxampsmalldisk} show, the apocenter/pericenter flux ratio
becomes insensitive to dust properties at millimeter wavelengths. For
a fixed disk mass, changes in $q$ and $\beta$ mostly affect the flux
contributed by the smallest particles: increasing $q$ increases the
number of very small particles, while increasing $\beta$ decreases the
flux reradiated at wavelengths larger than the particle size. However,
for the longest wavelengths, $\lambda\gg s$, the flux from small
particles becomes negligible. To lowest order the millimeter
apocenter/pericenter flux ratio therefore depends only on the
apocenter/pericenter temperature and density contrasts. The largest
bodies radiate efficiently and have effective temperature
\begin{equation}
  T\sim T_* (R_*/r)^{1/2}\propto 1\pm e/2
\end{equation}
where we assume the star is a blackbody with effective temperature
$T_*$ and radius $R_*$ and, in the last step, apply Eq.~\ref{eqn:r}
with $f=0$ (top sign) for pericenter and $f=\pi$ for apocenter (bottom
sign). Together with Eq.~\ref{eqn:ell} evaluated at $f=0$ and $\pi$,
this gives
\begin{equation}
  \mathrm{apocenter/pericenter\;flux\;ratio}
  \simeq \frac{1-e/2}{1+e/2}\frac{1+e}{1-e} \simeq 1+e
\end{equation}
where, as before, we take only lowest order terms in the eccentricity.
The millimeter apocenter/pericenter flux ratio thus provides a direct
estimate of the disk eccentricity.

Some systems, like $\epsilon$ Eridani, no doubt contain additional
structures that will complicate interpretation of their images. However,
with this new understanding of apocenter glow and its wavelength dependence,
we can begin future studies of debris disk images pointed in the
right direction.

\acknowledgments MP was supported by NSERC funds and by an NPP
fellowship at Goddard Space Flight Center administered by ORAU through
a contract with NASA. MP and MJK were partially supported by NASA
grant NNX15AK23G.  We thank an anonymous referee for knowledgeable
comments that improved our writeup. MP thanks Yoram Lithwick and
Yanqin Wu for helpful conversations, and Bok Tower Gardens for their
warm hospitality in the later stages of writing.

\end{document}